\documentclass[twocolumn,showpacs,preprintnumbers,amsmath,amssymb,prl]{revtex4}

 \def\XXint#1#2#3{{\setbox0=\hbox{$#1{#2#3}{\int}$}
     \vcenter{\hbox{$#2#3$}}\kern-.5\wd0}}

\def\fakebold#1{\relax\ifvmode\leavevmode\fi%
\ifmmode%
\setbox0=\hbox{$#1$}%
\else%
\setbox0=\hbox{#1}%
\fi%
\kern-.02em\copy0 \kern-\wd0%
\kern .04em\copy0 \kern-\wd0%
\kern-.0125em\raise.02em\box0%
}%

\usepackage{graphicx} \usepackage{dcolumn} \usepackage{bm}

\begin{document}



\title{Nonlinear 2D Spin Susceptibility in a Finite Magnetic Field:
  Spin-Polarization Dependence}

\author{Ying Zhang} 
\author{S. Das Sarma} 
\affiliation{Condensed
  Matter Theory Center, Department of Physics, University of Maryland,
  College Park, MD 20742-4111}

\date{\today}

\begin{abstract}

By theoretically calculating the interacting spin susceptibility of a
two dimensional electron system in the presence of finite
spin-polarization, we show that the extensively employed technique of
measuring the 2D spin susceptibility by linear extrapolation to
zero-field from the finite-field experimental data is theoretically
unjustified due to the strong nonlinear magnetic field dependence of
the interacting susceptibility.  Our work compellingly establishes
that much of the prevailing interpretation of the 2D susceptibility
measurements is incorrect, and in general the 2D interacting
susceptibility cannot be extracted from the critical magnetic field
for full spin polarization, as is routinely done experimentally.

\end{abstract}


\maketitle


The spin susceptibility, also called the Pauli susceptibility for the
non-interacting case, is a fundamental property of great significance
in condensed matter physics. For example, its behavior (e.g.
temperature dependence) could distinguish between Fermi and non-Fermi
liquids. The electron interaction induced density dependent
enhancement of spin susceptibility is a key signature of many body
effects in interacting Fermi liquids, which has been extensively
studied during the last fifty years~\cite{gellman, galitski, rice}. In
fact, the magnetic susceptibility of an itinerant electron system is
one of the key (as well as most-studied) thermodynamic properties of
metallic systems. In this Letter, we show theoretically that the
metallic magnetic susceptibility could depend rather strongly (and
non-trivially) on the spin polarization of the system, and such a
nonlinear polarization (or equivalently magnetic field) dependent spin
susceptibility could have profound effects on the interpretation of
many recent experimental measurements~\cite{vitkalov, shashkin0,
  tutuc, pudalov, zhu, shashkin} of 2D magnetic susceptibility in
confined semiconductor structures. In fact, we believe that our
theoretical work invalidates most of the recent interpretations of the
2D spin susceptibility measurements, particularly at lower carrier
densities and higher fields where the nonlinear effects are strong. We
emphasize that the spin-polarization (or the nonlinear field)
dependence of the magnetic susceptibility is {\em purely} an
interaction effect -- a strictly 2D noninteracting system has only the
usual linear free electron Pauli spin susceptibility.

The key theoretical idea introduced in this work is the observation,
almost obvious on hindsight (but routinely ignored in the extensive
recent experimental literature on the 2D susceptibility measurement),
that in a finite magnetic field B the net spin polarization of an
interacting 2D system is manifestly nonlinear in B, unlike the
corresponding linear noninteracting Pauli susceptibility situation.
This nonlinearity makes the experimental extraction of the interacting
2D susceptibility from a linear extrapolation of the finite-field
spin-polarized data to the zero-field limit, as is often done,
theoretically unjustified.

The specific relevance of our theoretical nonlinear susceptibility to
2D electron systems in semiconductor structures arises from the
particular experimental methods, involving the application of an
external magnetic field to spin-polarize the 2D system, typically used
to measure the 2D spin susceptibility~\cite{vitkalov, shashkin0,
  tutuc, pudalov, zhu, shashkin}. In one technique, a tilted magnetic
field, with components both parallel and perpendicular to the 2D
layer, is used, and the coincidence of the spin-split Zeeman levels
with the orbitally quantized Landau levels as manifested in the SdH
oscillations of the 2D magnetoresistence is used to obtain the Zeeman
energy and hence the susceptibility. In the other method, only an
applied parallel magnetic field is used to fully spin-polarize the 2D
system, and the observed kink in the magnetoresistence as a function
of the applied field is identified as the saturation field $B_c$ to
completely polarize the system, leading to the measured magnetic
susceptibility. We find that the strong nonlinear dependence of the
interacting 2D susceptibility on the applied magnetic field makes it
essentially impossible to extract the susceptibility from a
measurement of $B_c$, and some of the controversial conclusions in the
literature about the low-density behavior of the 2D susceptibility may
have arisen from $B_c$-based measurements. We note that both
experimental techniques involve spin-polarizing the 2D system, and
only when this spin-polarization is rather small in magnitude, the
susceptibility measurement is sensible.

For absolute theoretical clarity, we consider only the strict 2D limit
neglecting the quasi-2D layer thickness effect completely since the
finite layer thickness brings in the nonessential complications of the
parallel field induced magneto-orbital coupling~\cite{dassarma,
  tutuc2} already at the noninteracting level, leading to a rather
complex variation of the 2D susceptibility (due to the parallel
field-induced magneto-orbital coupling for motion perpendicular to the
2D layer) with the carrier density and the applied field, most
particularly at low (high) 2D densities (magnetic fields) when the
field-induced magnetic length is comparable to the finite layer
thickness. Since this is a conceptually simple (but numerically
intricate) one-electron band-structure effect, completely independent
of the many-body nonlinear effect of interest to us, we leave this
out, considering only the strict 2D theoretical limit where the
magneto-orbital coupling is, buy definition, absent. We neglect
thermal effects also, concentrating on $T=0$, in order to focus
entirely on the nonlinearity in the susceptibility.

A naive quasi-particle picture to determine the spin-polarization
$\zeta = (n_\uparrow - n_\downarrow)/n$ (where
$n_{\uparrow(\downarrow)}$ is the spin up (down) electron density and
$n = n_\uparrow + n_\downarrow$ is the total electron density) of the
2D electron system in an applied magnetic field $B$, is to separate
the spin-up quasiparticles and spin-down quasiparticles, and to use a
simple relation $E_{F\uparrow}^* - \mu_B B = E^*_{F\downarrow} + \mu_B
B$, where $E_{F\uparrow(\downarrow)}^*$ is the renormalized Fermi
energy for the spin up (down) quasiparticles, which is dependent on
the up (down) Fermi wavevecter $k_{F\uparrow} = k_F \sqrt{1+\zeta}$
($k_{F\downarrow} = k_F \sqrt{1-\zeta}$) with $k_F$ being the Fermi
wavevecter in the unpolarized state. Through this relation one can
determine $\zeta$, and then obtain the susceptibility. This naive
picture is suitable for deriving the zero-field susceptibility in the
limit $\zeta$ (or $B$) $\to 0$, and also for all fields in the
noninteracting electron model, but for the interacting system and at
finite fields, this simple relation does not hold. A more complete
theoretical treatment is then needed in considering the finite field
situation when eventually at some density dependent critical field
$B_c(n)$, the 2D system will undergo a first order transition to a
fully spin-polarized system. (At finite temperature, this first order
transition will be rounded, but the basic physics remains the same.)

We study the magnetization by calculating the total energy per
particle of the 2D system as a function of density, spin-polarization,
and magnetic field within the ring diagram
approximation~\cite{rajagopal, sus} which is exact at high density. In
an applied magnetic field $B$, the polarization $\zeta^*$ which
minimizes the energy then corresponds to the magnetization of the
system. The total energy per particle of the system can be written as
$E(r_s, \zeta, B) = E_K(r_s, \zeta) + E_Z(B) + E_{\rm C}(r_s, \zeta)$
where $E_K$ is the kinetic energy, $E_Z$ is the Zeeman energy due to
the finite magnetic field, and $E_{\rm C}$ is the interaction
(Coulomb) energy calculated within the many-body ring diagram
approximation. It is useful to mention here that the 2D spin
polarization properties (but not the nonlinear aspects of importance
in our work) have been theoretically studied with numerical quantum
monte Carlo techniques~\cite{attaccalite} which are in principle more
sophisticated than our analytic many-body approximation, but the
essential qualitative features (i.e.  the nonlinearlity in the
magnetic field) that are relevant for the present purpose are already
present in our ring-diagram calculation which becomes exact in the
high-density limit. We have used the notation of the interaction
parameter $r_s$, the so-called Wigner-Seitz radius, which is the
dimensionless inter-particle separation measured in the units of the
effective Bohr radius $a_B$: $r_s = (\pi n)^{-1/2}/a_B$.
It is easy to obtain
\begin{eqnarray}
\label{eq:EK}
E_K &=& {1 \over 2} ({k_{F\uparrow}^2 \over 2 m} {n_\uparrow \over n}
+ {k_{F\downarrow}^2 \over 2 m} {n_\downarrow \over n}) =
{1 + \zeta^2 \over 4 \alpha^2 r_s^2} (m a_B^2)^{-1}, \nonumber \\
\label{eq:EZ}
E_Z &=& - \mu_B B {n_\uparrow \over n} 
+  \mu_B B {n_\downarrow \over n} 
= - \mu_B B \zeta,
\end{eqnarray}
where $m$ is the electron mass, $\alpha = \sqrt{1/2}$, $\mu_B$ is the
electron magnetic moment (i.e. the Bohr magneton). The Coulomb energy
can be written as $E_{\rm C} = E_{\rm ex} + (E_{\rm C} - E_{\rm ex})$
where $E_{\rm ex}$, the exchange energy, can be written as
\begin{eqnarray}
\label{eq:Eex}
E_{\rm ex} = - {2  \over 3 \pi \alpha r_s} 
[(1 + \zeta)^{3/2} + (1 - \zeta)^{3/2}]
(m a_B^2)^{-1}.
\end{eqnarray}
The rest, the correlation energy, is then
\begin{eqnarray}
\label{eq:EC_Eex}
E_{\rm C} - E_{\rm ex}
= \int {d^2 q d \omega \over 2 n (2 \pi)^3} 
\left[ \ln(\varepsilon({\bf q}, i \omega)) 
- \varepsilon({\bf q}, i \omega) + 1 \right]
\nonumber \\
= {2 (ma_B^2)^{-1} \over \alpha^4 \pi r_s^2} 
\int_0^\infty x dx dz
[\ln \varepsilon(x, iz) - \varepsilon(x,iz) + 1]
\end{eqnarray}
where $\varepsilon({\bf q}, i \omega)$ is the dynamic dielectric
function~\cite{rice,sus}.

\begin{figure}[htbp]
\centering \includegraphics[width=2.5in]{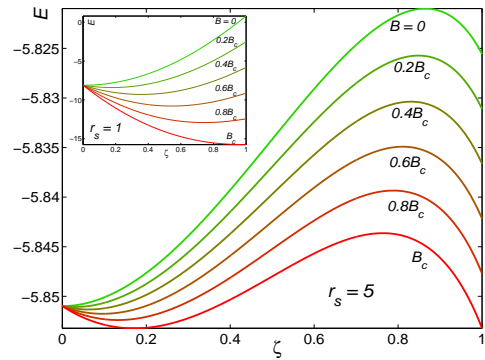}
  \caption{(Color online.) Calculated energy $E$ (in arbitrary units) 
    per particle as a function of spin polarization $\zeta$ in an
    applied magnetic field $B$ ranging from $0$ to $B_c$ with steps
    $0.2 B_c$ for $r_s = 5$ 2D electron system. (Note that $B_c$ is a
    function of $r_s$.)  Inset: the corresponding $r_s = 1$ results.}
  \label{fig:E}
\end{figure}

\begin{figure}[htbp]
\centering \includegraphics[width=2.5in]{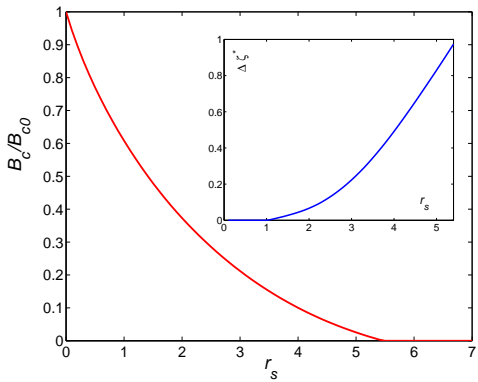}
  \caption{(Color online.) Calculated full polarization critical 
    magnetic field $B_c$ as a function of $r_s$ in units of the
    corresponding non-interacting value $B_{c0}$. Inset: the
    discontinuous jump of spin polarization $\zeta^*$ at $B_c$}
  \label{fig:Bc}
\end{figure}

In Fig.~\ref{fig:E} we present the energy per particle $E$ as a
function of spin polarization $\zeta$ in different applied magnetic
field $B$. As we can see from Fig.~\ref{fig:E}, for small enough $r_s$
($r_s < r_s^* \sim 5.5$, the value of which is obvious from
Fig.~\ref{fig:Bc}), the system prefers zero spin polarization at
$B=0$.  As $B$ increases, the energy curve shifts down while the
minimum energy corresponds to a non-zero spin polarization $\zeta^*$.
When $B$ increases to $B_c$, there exist two $\zeta^*$ values which
minimize the energy. For example, in $r_s = 5$ case as shown in
Fig.~\ref{fig:E}, when $B=B_c$ one energy minimum corresponds to
$\zeta^* = 0.15$ and the other corresponds to $\zeta^* = 1$. For all
$B>B_c$ cases, the energy minimum always corresponds to $\zeta^* = 1$.
This means that as $B$ increases from just below to just above $B_c$,
$\zeta^*$ suddenly jumps by $\Delta \zeta^*$ ($\Delta \zeta^* = 0.85$
in $r_s = 5$ case) from a value less than $1$ ($0.15$ in $r_s = 5$
case) to $1$, and the system undergoes a first order transition to a
spin-polarized state.  Note from the inset of Fig.~\ref{fig:E} that
when $r_s$ is small, the downward trend of the energy curve at large
$\zeta$ value is not strong, and it seems at $B = B_c$, there is only
one energy minimum. A closer inspection of the energy curve yields the
fact that there actually exists two minima, only too close to each
other to be noticed in the figure. Therefore the spin polarization
transition in the presence of the finite field $B$ is still first
order even for a small $r_s$ system, only with a small $\Delta
\zeta^*$ value. The important point to note here is that the
field-induced transition to the full spin-polarization at $B=B_c$ is
always first-order, accompanied by a finite discontinuity in the spin
polarization. 

The ground state energy per particle as a function of $B$ and $r_s$ is
an important result, from which other physical quantities can be
derived. For example, the critical polarization magnetic field $B_c$,
which is a function of $r_s$, can be determined through the above
procedure for each $r_s$ value. Using the polarization magnetic field
for non-interacting 2D electron gas system $B_{c0} = E_F / \mu_B$ as
the unit, we plot the $B_c$ for the interacting 2D electron system as
a function of $r_s$ in Fig.~\ref{fig:Bc}.  From this figure we see
that $B_c$ decreases monotonically as $r_s$ increases, and that at
$r_s = r_s^* (\sim 5.5)$, $B_c$ decreases to zero, and the system is
spontaneously spin-polarized. This result confirms those of previous
theoretical calculations~\cite{rajagopal, sus} in the ring diagram
approximation. In the inset of Fig.~\ref{fig:Bc} we show the discrete
jump of the spin polarization at $B = B_c$ as a function of $r_s$. We
emphasize that the exact value of $r_s^* (\approx 5.5)$ here depends
on the model and the approximation scheme, and is much
larger~\cite{sus} for realistic quasi-2D systems. Also at finite $T$,
the abrupt discontinuity is smoothened somewhat.

\begin{figure}[htbp]
\centering \includegraphics[width=2.5in]{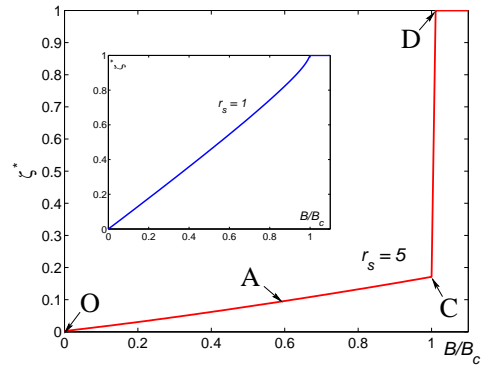}
  \caption{(Color online.) Calculated spin polarization as a function
    of magnetic field $B$ for $r_s = 5$. Inset: the corresponding $r_s
    = 1$ results. The relevance of O, A, C, D in defining various
    susceptibility are discussed in the text.}
  \label{fig:z}
\end{figure}

\begin{figure}[htbp]
\centering \includegraphics[width=2.5in]{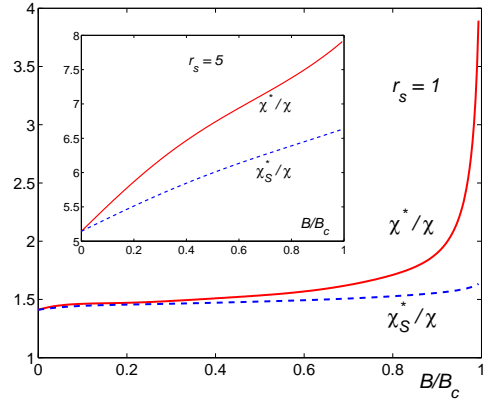}
  \caption{(Color online.) Calculated spin susceptibility $\chi^*$
    (red solid curves) and semi-linear spin susceptibility $\chi^*_S$
    (blue dashed curves) as a function of magnetic field $B$ for $r_s
    = 1$ 2D electron system. (The tilted field measurements
    essentially obtain $\chi^*_S$.) Inset: the corresponding $r_s = 5$
    results.  }
  \label{fig:chiB}
\end{figure}

From the ground state energy we are able to determine the
magnetization curve $\zeta^*(B)$ (Fig.~\ref{fig:z}), from which we
notice that the magnetization increases as a convex function of $B$
(the convexity is seen clearly in the increasing of the susceptibility
shown in Fig.~\ref{fig:chiB}), and experiences a discrete jump at
$B=B_c$. For $B > B_c$, the system remains fully polarized ($\zeta^*
=1$). As mentioned, the magnetization jump in small $r_s$ system is
less pronounced.

The nonlinear spin susceptibility $\chi^* = n (d \zeta^* /d B)$ can be
derived from magnetization $\zeta^*$ shown in Fig.~\ref{fig:z}. Since
the magnetization curve has a jump at $B = B_c$, the spin
susceptibility $\chi^*$ is only meaningful for magnetic field within
the range of $0 \le B < B_c$. In Fig.~\ref{fig:chiB} we present
calculated spin susceptibility (using the non-interacting Pauli
susceptibility $\chi$ as the unit) as a function of $B$ for two
different $r_s$ values: $r_s = 1$ and $5$. It is worth mentioning that
$\chi^*$ always increases with increasing $B$, i.e. the nonlinearity
of the interacting 2D susceptibility is a monotonically increasing
function of $B$ in the $0<B<B_c$ range. The quantitative behavior of
nonlinear $\chi^*(r_s;B/B_c)$ is also a strong function of $r_s$, as
one can see by comparing the main figure and the inset in
Fig.~\ref{fig:chiB}. The susceptibility remains finite for all $B$ up
to $B_c$, after which $\chi^*$ is not well-defined. (In
Fig.~\ref{fig:chiB} we also show the result for, what we call, the
{\em semi-linear spin susceptibility} $\chi^*_S$, which is related to
experimental studies of the susceptibility and is defined below.)

We have also calculated the zero-field susceptibility ($n
(d\zeta^*/dB)|_{B \to 0}$), finding precise agreement with our earlier
results~\cite{sus}. We emphasize, however, that the experimental
measurements~\cite{vitkalov, shashkin0, tutuc, pudalov, zhu, shashkin}
do not typically measure the nonlinear susceptibility shown in
Fig.~\ref{fig:chiB} or the zero-field susceptibility although most
experimental interpretations automatically (and as we show in this
Letter, incorrectly) assume that the experimentally measured
susceptibility is the usual zero-field linear susceptibility.

One experimental way to study the spin susceptibility is to obtain the
polarization field $B_c$ through magneto-resistance
measurements~\cite{vitkalov, shashkin0, tutuc, pudalov, shashkin}, and
then obtain the ``spin susceptibility'' from $B_c$ using the
non-interacting formula. In fact this is not really the spin
susceptibility $\chi^* = n (d \zeta/d B)|_{B = 0}$, but a different
quantity which we call the {\em linear spin susceptibility}: $\chi^*_L
= n/B_c$. In Fig.~\ref{fig:z}, the susceptibility $\chi^*$ is
represented by the derivative of the curve at point `O', while the
{\em linear spin susceptibility} $\chi^*_L$ is represented by the
slope of line `OD'. These two quantities $\chi^* (B=0)$ and $\chi^*_L$
(measured experimentally from the slope of the line `OD' in
Fig.~\ref{fig:z}) are certainly very different from each other,
especially at larger $r_s$ values. We also note that the real critical
field $B_c(D)$ corresponding to the point `D' is much smaller than the
extrapolated line `OC' would indicate! In particular, $\chi^*_L$ would
always be much larger than $\chi^*(B \to 0)$, and the experimental
conclusion based on the measurement of $B_c$ is simply incorrect. It
should be noted in this context that the semi-linear susceptibility
$\chi^*_S$ (shown in Fig.~\ref{fig:chiB} and discussed below) is
always smaller in magnitude than $\chi^*$, and therefore in general,
$\chi^*_L > \chi^*_S$.

Another experimental method (the tilted field method) to study the
susceptibility is by matching Landau levels and Zeeman energy
levels~\cite{zhu}.  The experimental detail boils down to measuring,
what we call, the {\em semi-linear spin susceptibility} $\chi^*_S (B)
= n \zeta^*(B)/B$, shown in Fig~\ref{fig:chiB}.  The easiest way to
describe this quantity is by examining Fig.~\ref{fig:z}. The {\em
  semi-linear spin susceptibility} $\chi^*_S(B)$ at point $A$ is
represented by the slope of line `OA', while the susceptibility
$\chi^*(B)$ is represented by the derivative of the magnetization
curve at point `A'. Of course these two quantities are different,
especially in a large magnetic field, as shown in Fig.~\ref{fig:chiB}.
However, the experimental measurement of this {\em semi-linear spin
  susceptibility} $\chi^*_S$ is still reasonably meaningful in the
following ways. One is that for $B=0$, $\chi^*_S$ and $\chi^*$
coincide with each other as shown in Fig.~\ref{fig:chiB}, and
therefore {\em theoretically speaking}, this measurement~\cite{zhu}
should be able to capture the true behavior of the zero-field
susceptibility. Another meaningful aspect of this experiment is that
the measurement~\cite{zhu} shows that,as $B$ increases, $\chi^*_S$
also increases~\cite{zhu}, which suggests that the magnetization curve
is convex even though $\chi^*_S$ and $\chi^*$ are different. This
observation agrees with our theoretical findings. We therefore
conclude that the tilted field measurement leading to $\chi^*_S$ is
reasonable (but still far from perfect) for measuring the 2D
susceptibility for $B<B_c$, whereas the susceptibility $\chi^*_L$
(extracted from the measurement of $B_c$) is not particularly
meaningful.

In conclusion, we have calculated the nonlinear magnetization and spin
susceptibility as a function of magnetic field and density for 2D
electron systems with long-ranged Coulomb interaction in an applied
magnetic field. We find that most measurements of 2D spin
susceptibility are incorrect because they do not incorporate the
magnetic field-induced nonlinearity.  Because of our neglect of sample
details (e.g. finite width effects), our general theory is not
directly comparable to the existing experimental data in any
particular system, but our work establishes that any experiment in a
finite magnetic field, cannot provide a meaningful measurement of the
2D susceptibility, except at the lowest fields and highest densities
(i.e. for $B \ll B_c$) where our predicted nonlinear effects are
quantitatively small. In particular, we show convincingly that an
experimental measurement of $B_c$ (e.g.  the parallel field
magneto-transport data) most certainly does {\em not} provide a value
for the zero-field interacting 2D susceptibility as has been
uncritically assumed in most earlier works whereas the tilted field
measurements, particularly in thin 2D samples at low magnetic fields,
provide an approximate measurement of the susceptibility. Finally, we
note that finite temperature effects would smoothen the discontinuity
(at $B_c$) in the magnetization since there will be some finite
thermal population of both spin up/down bands, but the same physics
will apply qualitatively at low temperatures.

This work is supported by ONR, NSF, and LPS.


\end{document}